\documentclass[twocolumn]{aastex63}

\usepackage{amsmath}
\usepackage{amssymb}
\usepackage{amsthm}
\usepackage{float}
\usepackage{natbib,aasdefs,url,bm}
\usepackage{graphicx}
\usepackage{color}
\usepackage{mathtools, epsfig, lipsum}
\usepackage{subfigure}
\usepackage{textcomp}

\usepackage{multirow,booktabs,colortbl,tabularx}
\usepackage{enumitem}
\usepackage{booktabs}
\usepackage{hyperref}
\usepackage{lineno}
\usepackage{silence}
\usepackage{CJK}
\shorttitle{High-energy Neutrino Source Cross-correlations with Nearest-neighbor Distributions}
\shortauthors{Zhou, Cisewski-Kehe, Fang, \&Banerjee}

\begin{document}
\begin{CJK*}{UTF8}{gbsn}
\title{High-energy Neutrino Source Cross-correlations with Nearest-neighbor Distributions}

\author{Zhuoyang Zhou (周卓扬)}
\altaffiliation{Current affiliation: Department of Statistics \& Data Science, Carnegie Mellon University, Pittsburgh, PA, USA}
\affiliation{Department of Statistics, University of Wisconsin-Madison, Madison, WI, USA}
\affiliation{Department of Statistics \& Data Science, Carnegie Mellon University, Pittsburgh, PA, USA}

\author{Jessi Cisewski-Kehe}
\affiliation{Department of Statistics, University of Wisconsin-Madison, Madison, WI, USA}

\author{Ke Fang}
\affiliation{Department of Physics, Wisconsin IceCube Particle Astrophysics Center, University of Wisconsin-Madison, Madison, WI, USA}

\author{Arka Banerjee}
\affiliation{Department of Physics, Indian Institute of Science Education and Research, Pashan, Pune, India}

\begin{abstract}
 The astrophysical origins of the majority of the IceCube neutrinos remain unknown. Effectively characterizing the spatial distribution of the neutrino samples and associating the events with astrophysical source catalogs can be challenging given the large atmospheric neutrino background and underlying non-Gaussian spatial features in the neutrino and source samples. In this paper, we investigate a framework for identifying and statistically evaluating the cross-correlations between IceCube data and an astrophysical source catalog based on the $k$-Nearest Neighbor Cumulative Distribution Functions ($k$NN-CDFs). We propose a maximum likelihood estimation procedure for inferring the true proportions of astrophysical neutrinos in the point-source data. We conduct a statistical power analysis of an associated likelihood ratio test with estimations of its sensitivity and discovery potential with synthetic neutrino data samples and a WISE-2MASS galaxy sample. We apply the method to IceCube's public ten-year point-source data and find no statistically significant evidence for spatial cross-correlations with the selected galaxy sample. We discuss possible extensions to the current method and explore the method's potential to identify the cross-correlation signals in data sets with different sample sizes. 

\end{abstract}


\section{Introduction}
The existence of extraterrestrial astrophysical neutrino emission has been confirmed by the IceCube Neutrino Observatory \citep{Aartsen2013}. More recently, evidence of high-energy neutrino emission from individual sources has been established \citep{IceCube:2018cha, IceCube:2019cia,Abbasi2022}. The astrophysical origins of the majority of detected neutrino events, however, are still largely unknown.

The most commonly used neutrino source search method involves computing the probability of each neutrino coming from each candidate in a catalog or each pixel in a source template \citep{Braun:2008bg}. This method has successfully led to the detection of neutrinos from NGC~1068 \citep{Abbasi2022} and the Galactic Plane \citep{IceCube:2023ame}. It has also been applied to study the association of neutrinos with nearby galaxies \citep{IceCube:2019yml}. Another method is to search for the cross-correlation between the spatial distributions of neutrino and source samples \citep{Fang2020, 2021A&A...652A..41U, 2022PhRvD.105l3035C, 2023ApJ...951...83N,2023arXiv230803978G, Ouellette:2024ggl}. As the cross-correlation method only relies on the assumption that the neutrino sources trace the same underlying field as the chosen catalog, it has the advantage of not requiring a perfect catalog that contains the actual neutrino sources. 

So far the cross-correlation studies are based on the two-point correlation function (2PCF). Such a correlation analysis could be limiting since 2PCF cannot give a comprehensive characterization of non-Gaussian spatial clustering that is present in the distribution of matter at low redshifts and on small spatial separation scales ($\lesssim 40 \,h^{-1}{\rm Mpc}$). One possible solution is to consider higher $N$-point correlation functions. However, the associated computational cost will be much higher as the computational complexity increases exponentially with $N$, which makes it impractical to measure on large datasets \citep{Banerjee2021a}. Given the large atmospheric background in the full IceCube point source data, a more sensitive measure of spatial clustering with satisfying computational cost becomes necessary.

To this end, we adopt the $k$-Nearest Neighbor Cumulative Distribution Functions \citep{Banerjee2021a}, $k$NN-CDFs, in our cross-correlation study. For a group of data points that fill a volume and a specific $k$, the $k$NN-CDFs are simply defined as the empirical cumulative distribution function of the associated $k$-nearest neighbor distances, which can be estimated through interpolation on the sorted $k$NN distances. In a three-dimensional space with an Euclidean metric, the value of the $k$NN-CDFs at a scale $r$ can be interpreted as the estimated probability of finding at least $k$ data points that are enclosed in a sphere with radius $r$. Based on the $k$NN-CDFs, a measure of the similarities of spatial clustering of two data samples called joint $k$NN-CDFs and an associated $k$NN-CDFs cross-correlation is available \citep{Banerjee2021b}. 

As mathematically shown in \citet{Banerjee2021a}, the $k$NN-CDFs are related to the $N$-point correlation functions measured on the same group of data points but have a reduced computational burden through the use of $k$-d trees that can be constructed with $N\log N$ operations and a $k$NN searching in $\log N$ time for each point. The $k$NN-CDFs have demonstrated more significant detection of spatial clustering in sparse datasets such as the 1000 richest redMaPPer clusters in the SDSS DR8 redMaPPer clusters than the two-point correlation function \citep{Wang2022}. 

In this work, we adapted the $k$NN-CDF method to accommodate the properties of IceCube point source data, where only the equatorial coordinates are available for each point source without the depth (redshift) information. We developed the framework to compute the $k$NN-CDFs cross-correlations and a test of significance of such correlation between IceCube 10-year point source data \citep{abbasi2021icecube, publicData} and a galaxy catalog based on the Wide-Field Infrared Survey Explorer (WISE; \citealp{Wright2010}) and the 2-Micron All-Sky Survey (2MASS; \citealp{2006AJ....131.1163S}) infrared databases. By estimating the sensitivity and discovery potential on the synthetic neutrino datasets generated based on the public ten-year data, we have shown that the methodology is capable of identifying a highly significant cross-correlation when the neutrino events and the galaxy sample both trace the same underlying large-scale density field.

The layout of the paper is as follows: we introduce the backgrounds of $k$NN-CDFs and related summary statistics in Section~\ref{sec:backgrounds}. The $k$NN-CDFs-related computations are introduced in Section~\ref{sec:methodology}. In Section~\ref{sec:setup}, we summarize the data engineering for neutrino and galaxy samples that are used in the analysis. Then we present the main findings based on synthetic samples and the IceCube public ten-year data in Section~\ref{sec:experiments} and Section~\ref{sec:results}, respectively. A discussion and concluding remarks are provided in Section~\ref{sec:discussion}.

\section{Background}\label{sec:backgrounds}
The goal of the $k$NN-CDFs is to offer a more sensitive and comprehensive characterization of the density field from which the data samples (discrete tracers) are drawn. Thus, a preliminary assumption is that the observed data points are indeed the discrete tracers of their underlying density field, and the degrees of spatial clustering of the data points directly reflect their underlying density fluctuations. Note that the $k$NN-CDF can be viewed as a generalization of the G-function in the statistics literature, where the G-function uses $k=1$; the G-function and other spatial point process functions are discussed in \cite{baddeley2015spatial} and \cite{diggle2013statistical} 

\subsection{\texorpdfstring{$k$NN-CDFs}{}}\label{sec:kNN}
We start by defining the $k$NN-CDFs and leave the detailed computation in the methodology Section \ref{sec:methodology}. Assume there is a group of data points distributed over a three-dimensional volume $V$, for a specific $k$, the value of $k$NN-CDF at distance $r$, denoted as $\text{CDF}_k(r)$, can be defined as the probability of finding at least $k$ data points at a distance less than $r$ for a random point in the same volume. With the usual Euclidean metric, we have the following definition: 
\begin{linenomath*}
\begin{equation} \label{eq:1}
    \text{CDF}_k(r) = P_{>k-1|V}|_{V=\frac{4}{3}\pi r^3},
\end{equation}
\end{linenomath*}
which can be interpreted as the probability of finding spheres of radius $r$ that enclose at least $k$ data points. This gives the $k$NN-CDFs an intuitive interpretation as measures of the degree of spatial clustering at different scales with these ``probing spheres'' on the data sample. 

Notice that $k$NN-CDFs can also be defined on two-dimensional surfaces and other Riemannian manifolds with a corresponding metric, such as the surface of a unit sphere with a metric defined as the great circle distance. This is crucial for the application in the context of IceCube point source data since only the equatorial coordinates are available for each point source without the depth information or redshift. Thus, a projection onto a spherical surface is needed.

\subsection{Joint \texorpdfstring{$k$NN-CDFs}{}}\label{sec:joint kNN}
After defining the $k$NN-CDFs on a single data sample, it can be extended to a joint function called the joint $k$NN-CDFs based on two data samples. Similar to the definition of $k$NN-CDFs, for specific $k$ (e.g., $k_1=k_2=k$), the value of the joint $k$NN-CDF at $r$, denoted as $\text{CDF}_{k_1,k_2}(r)$, can be defined as the probability of finding a sphere with radius $r$ that contain at least $k$ data points in both data samples:
\begin{linenomath*}
  \begin{equation} \label{eq:joint}
    \text{CDF}_{k_1,k_2}(r) = P(>k_1-1, >k_2-1|V).
\end{equation} 
\end{linenomath*}

As shown in \citet{Banerjee2021b}, there is a relationship between the joint $k$NN-CDFs and the connected $N$-point correlation functions such that the combined joint $k$NN-CDFs contain the same information as the $N$-point correlation functions that can be measured on the two density fields from which the two data samples are drawn. In other words, the joint $k$NN-CDFs effectively measure the correlation of the two data samples in terms of spatial clustering. 

\begin{figure}
\includegraphics[width= 1 \linewidth] {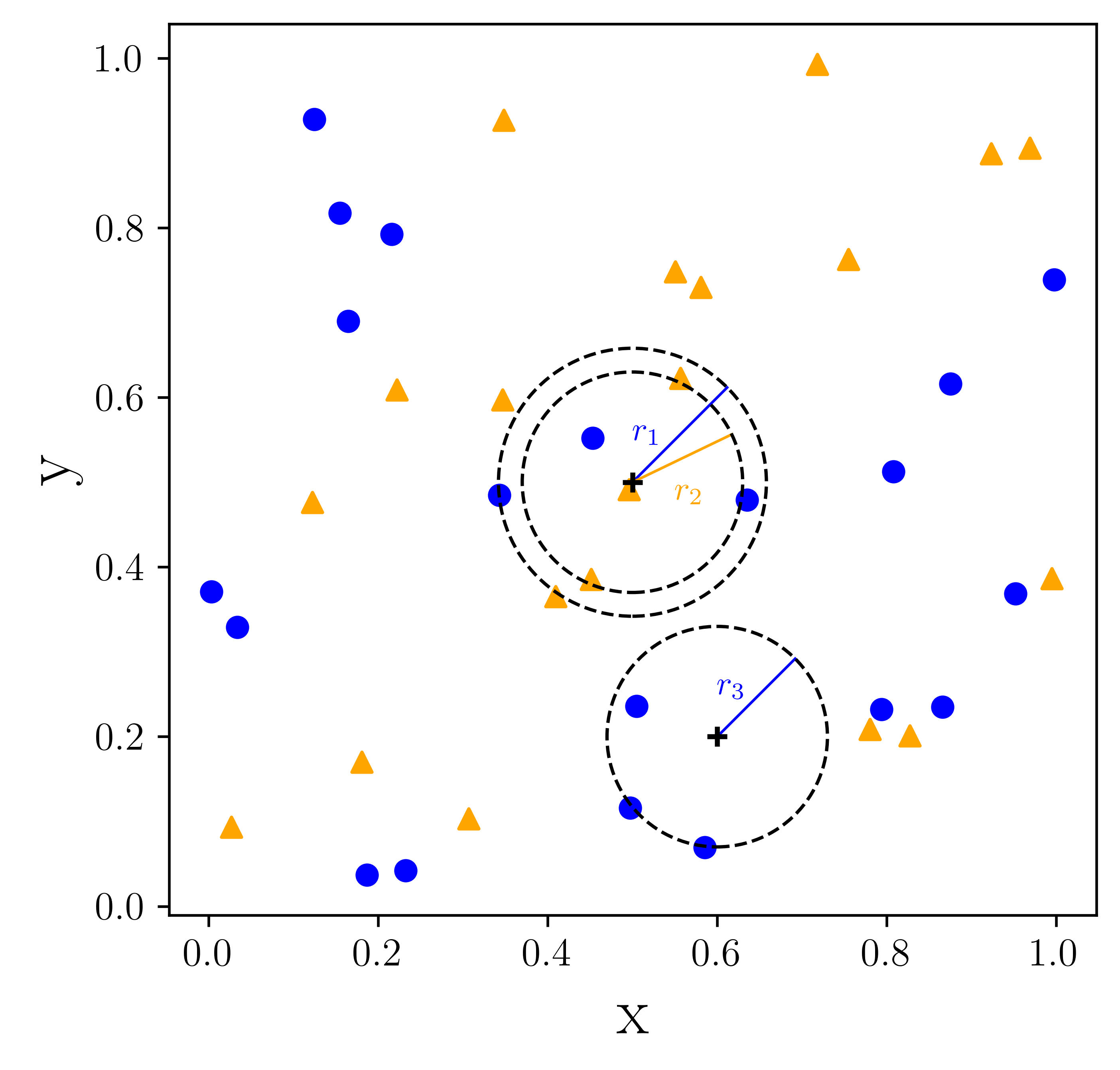} 
\caption{\label{fig:knn} Illustration of the $k$NN-CDFs and joint $k$NN-CDFs computations for $k=3$ with the Euclidean metric. The orange triangles and blue circles represent the source catalog (IceCube neutrinos in our case) and tracers (such as the WISE-2MASS samples), respectively. The two black pluses are reference points to demonstrate the calculation of $3$NN-CDFs and joint $3$NN-CDFs, respectively. }
\end{figure}

Before discussing the $k$NN-CDFs cross-correlation, we provide a visual demonstration in Figure \ref{fig:knn} for a more intuitive understanding of the computation of the $k$NN-CDFs and the joint $k$NN-CDFs within a rectangular region using the usual Euclidean metric (this can be other surfaces or volumes with an appropriate metric in practice). The black pluses are reference points generated from a homogeneous spatial Poisson process. The orange and blue points, respectively, represent the source catalog (IceCube neutrinos in our case) and tracers (such as the WISE-2MASS samples). In practice, many more reference points (black pluses) are generated. We only consider two for illustration purposes. The reference point at the bottom of Figure~\ref{fig:knn} illustrates the 3NN-CDF computation. Its radius $r_3$ represents one  3NN distance to the blue points. The 3NN-CDF of the blue points are estimated with the empirical 3NN-CDF based on many such $r_3$ from different reference points. The reference point at the center illustrates the computation of the joint $3$NN-CDF. This reference point identifies two 3NN distances, $r_1$ and $r_2$, to its nearby orange and blue points, respectively. We adopt the larger of the two, namely $\max(r_1, r_2)=r_1$, as the joint distance. We apply such a procedure on all reference points and obtain the empirical CDF from the joint distances as the joint 3NN-CDF between the blue and orange points.

\subsection{\texorpdfstring{$k$NN-CDFs}{} Cross-correlation}\label{sec:knn-cc}
To isolate the effect of the clustering from two individual data samples (density fields), we can subtract the product of the individual $k$NN-CDFs that are measured from each dataset and obtain a measure of the $k$NN-CDFs cross-correlation of the two data samples:
\begin{linenomath*}
\begin{equation}\label{eq:3}
    \psi^{k_1,k_2}(r)=\text{CDF}_{k_1,k_2}(r)-\text{CDF}_{k_1}^{(1)}(r)\cdot\text{CDF}_{k_2}^{(2)}(r).
\end{equation}  
\end{linenomath*}

Higher positive values of the cross-correlation defined above imply a higher level of spatial correlation between the two density fields, while negative values indicate anti-correlations \citep{Banerjee2021b}.

Note that while $k$NN-CDFs and the $k$NN-CDFs cross-correlation can be applied on samples with different sample sizes by choosing $k_1\neq k_2$ in Equation \eqref{eq:3}. For simplicity, in this paper, we restrict our scope to cases when $k_1=k_2=k$ and require that the two data samples have roughly the same number of data points. To reach this setting, the WISE-2MASS galaxy sample is down-sampled to match the sample size of the neutrino catalog. A discussion of how to handle datasets with very different sizes is presented in Section~\ref{sec:discussion}.

\section{Methodology}\label{sec:methodology}
In this section, we outline how to compute the $k$NN-CDFs, joint $k$NN-CDFs, and the $k$NN-CDFs cross-correlation we defined in Section \ref{sec:backgrounds} respectively, specifically on the IceCube point-source data with adaptations of the original computations outlined in \cite{Banerjee2021a} and \cite{Banerjee2021b}. 

\subsection{Estimating \texorpdfstring{$k$NN-CDFs}{}}\label{sec:knn}
To estimate the $k$NN-CDFs, we first generate a group of reference points from a homogeneous spatial Poisson process that fills the volume (or surface) for both the IceCube point-source data and the galaxy catalog. In IceCube's context, since all data points lack depth information, we can treat them as data points distributed on the surface of a unit sphere with right ascensions (RAs) and declinations (Decs) in radians. The homogeneous spatial Poisson process is generated on the same unit sphere.

At a specific $k$, instead of performing $k$NN searching within the data sample, we measure all the $k$-nearest neighbor distances from the reference points to the IceCube point-source data or the selected tracers. To better characterize the two tails of the empirical CDF, we use more reference points than the data points. We can immediately estimate the $k$NN-CDFs by obtaining the empirical cumulative distribution function by using a 1-D linear interpolation on the sorted $k$NN distances. As noted by \citet{Banerjee2021a}, $k$-nearest neighbor searching can be done with high computational efficiency by constructing a $k$-d tree \citep{bentley1975multidimensional} on a total of $N$ data points with $NlogN$ complexity, and the nearest neighbor searching can be completed in $logN$ time for each reference point. 

In our case, instead of using a $k$-d tree and the Euclidean metric, the nearest neighbor searching is performed using the great-circle distances by constructing a ball-tree \citep{omohundro1989five} on the data points with the Scikit-learn's \texttt{sklearn.neighbors.BallTree} implementation \citep{JMLR:v12:pedregosa11a} using the haversine distance metric (based on great-circle distances). For the above implementation, the great-circle distance between two points $x,y$ in the geographic coordinate system is defined as:
\begin{linenomath*}
\begin{equation*}\label{eq:d}
      d(x,y) = 2r\arcsin{\sqrt{\sin^2({\frac{\Delta\phi}{2}}) +\cos{\phi_{x}}\cos{\phi_y}\sin^2({\frac{\Delta\lambda}{2}})}},
\end{equation*} 
\end{linenomath*}
where $r$ is the radius of the sphere ($r=1$ in our case) and $\Delta\phi=\phi_x-\phi_y$ and $\Delta\lambda=\lambda_x-\lambda_y$. The $\phi$ and $\lambda$ are the latitude and longitude (both in radians), respectively, of the two data points. We replace the latitudes and longitudes with an equatorial coordinate system using RA and Dec (J2000) to match the format of the IceCube public point-source data. The choice of a ball-tree is due to the fact that while the $k$-d tree partitions the entire space along the axis ($x$, $y$, and $z$ axis in three-dimensional Cartesian spaces), the ball-tree only partitions the manifolds on which the data points are located, and is more suitable when sources are all projected on a spherical surface.

\subsection{Estimating Joint \texorpdfstring{$k$NN-CDFs}{} and Cross-correlation}\label{sec: join and cc}
When considering two data samples at a specific $k$, each reference point is separately associated with two $k$NN distances, each from one of the datasets. To measure the joint term $\text{CDF}_{k_1,k_2}(r)$, for a reference point at a specific $k=k_1=k_2$, the associated $k$NN distances used to obtain the estimated CDF is the maximum of the distances measured from the two data samples. Similar to the interpretation of $k$NN-CDFs on a single dataset, such a joint term can be interpreted as the probability of finding spheres of radius $r$ that enclose at least $k$ data points from both datasets \citep{Banerjee2021a}. This effectively serves as a measure of the similarities in terms of spatial clustering of the two data samples.

For both the $k$NN-CDFs and joint $k$NN-CDFs, the $k$ corresponding empirical CDFs are estimated with linear interpolation using Python's \texttt{scipy.interpolate.interp1d} \citep{2020SciPy-NMeth} on the middle 90\% (between the 5th and 95th quantiles) of the sorted $k$NN distances to avoid over-extrapolating into both tails where there may be only a few data points. Then we generate 50 linearly separated bins (in radians) on the same distance range for all chosen $k$. Each $k$NN-CDF and joint $k$NN-CDF are represented by length-50 vectors, where each entry represents the value of the interpolating function at the corresponding radius. The final $k$NN-CDFs cross-correlations are calculated directly by plugging in the obtained quantities in Equation \eqref{eq:3}. We provide sample code to demonstrate our $k$NN-CDF and joint $k$NN-CDF calculations in the {\it Software} section at the end of Section \ref{sec:conclusion}.

\section{Data Samples and Analysis Setup}\label{sec:setup}
For a demonstration of our method, we perform a $k$NN-CDFs cross-correlation analysis using a neutrino sample and a galaxy catalog.  In this section, we describe the neutrino and galaxy samples used in the study and the procedures for synthetic events generation. 

\subsection{Galaxy and IceCube Neutrino Sample}\label{sec:galaxy}
We use the public all-sky point-source IceCube data from years 2008-2018 \citep{abbasi2021icecube}. To avoid the large atmospheric muon background, we select neutrino events from the northern sky with Dec $\delta>-5^\circ$. We mask the Galactic Plane between Galactic latitudes $|b|<10^\circ$. 

The WISE-2MASS galaxy catalog is constructed by combining the photometric data from the WISE \citep{Wright2010} and the 2MASS \citep{2006AJ....131.1163S} infrared databases following \citet{2015MNRAS.448.1305K} as in \citet{Fang2020}. 
As explained in Section~\ref{sec:knn-cc}, we down-sample the galaxy catalog by randomly selecting $n=451,953$ (about 17\% of the original data) galaxies to match the sample size of the neutrino data after the masking. 

\subsection{Synthetic Neutrino Samples}\label{sec:synthetic}
Synthetic neutrino samples are used to conduct statistical power analyses. Atmospheric events are generated by randomizing the RAs of the original data. To generate astrophysical neutrinos, we first randomly sample a fixed number of sources from the galaxy catalog. The sources are assumed to be equally bright and each follows a power-law energy spectrum $dN/dEdAdt = A_0\, (E / 10^3\,{\rm GeV})^{-2}$ between  $E_{\rm min} = 100 \, \rm GeV$ and $E_{\rm max} = 10^9 \,\rm GeV$. Then for each selected source, we generate events based on the effective areas and angular resolutions of the public point-source data. The number of events emitted by a source is a Poisson random number with the mean being the assumed source flux integrated over the detector effective area for a given source direction. Finally, we combine background-only events and astrophysical events from the selected galaxies to obtain a synthetic sample that contains a total of $N_{\rm ev}$ events and ${f}_{\text{astro}}$ fraction of astrophysical events. 

The flux normalization is fixed to $A_0 = 5\times{10}^{-16}\,\rm GeV^{-1}\,cm^{-2}\,s^{-1}$ in our default cases, which corresponds to a median of 5 events per source. The source number is determined by the total number of astrophysical events, $N_{\rm ev}\,f_{\rm astro}$, divided by the event per source. For example,  with ${f}_{\text{astro}} = 3.74\%$ and the default $A_0$, the number of sources selected to inject neutrinos is 3500. Our synthetic event generation code is provided in the {\it Software} section at the end of Section \ref{sec:conclusion}. 

\section{Likelihood and Sensitivity}\label{sec:experiments}
Below we define the likelihood function and an associated likelihood ratio test for the computed $k$NN-CDFs cross-correlation (referred to simply as ``cross-correlation'' in what follows). We verify the properties of the likelihood and estimate the sensitivity and discovery potential of the search. 

\subsection{Likelihood Function and Likelihood Ratio Test}\label{sec:likelihood}
Since the astrophysical and atmospheric neutrino come from unrelated sources with different density fields, there should not be a contribution to the correlation with the WISE-2MASS samples. Then the cross-correlation approximately satisfies:
\begin{linenomath*}
\begin{equation}\label{eq:approx}
  \boldsymbol{\psi^{(k)}_{\text{g,v}}}(r) = f_{\text{astro}}\cdot\boldsymbol{\psi^{(k)}_{\text{g,astro}}}(r) + (1-f_{\text{astro}})\cdot\boldsymbol{\psi^{(k)}_{\text{g,atm}}}(r),
\end{equation}   
\end{linenomath*}
where $\boldsymbol{\psi^{(k)}_{\text{g,v}}}(r)$ denotes the cross-correlation with $k=k_1=k_2$ between a neutrino sample of mixed astrophysical and atmospheric signals with a specific proportion of astrophysical signals, $f_{\text{astro}}$, and the galaxy sample. The $\boldsymbol{\psi^{(k)}_{\text{g,astro}}}(r)$ denotes the cross-correlation between a neutrino sample with only astrophysical ($f_{\text{astro}}=1$) events and the galaxy catalog, and $\boldsymbol{\psi^{(k)}_{\text{g,atm}}}(r)$ denotes the cross-correlation between a neutrino sample with only atmospheric ($f_{\text{astro}}=0$) events and the galaxy catalog. Notice that there should not be a spatial correlation between the galaxy sample and the atmospheric-only neutrinos. Hence, $\boldsymbol{\psi^{(k)}_{\text{g,atm}}}$ likely captures a measurement noise due to the spatial distribution of the atmospheric neutrinos and the declination dependence of the detector effective area, which can be considered as the benchmark against the detection of astrophysical neutrinos. The relationship in Equation \eqref{eq:approx} is verified with synthetic samples in Section \ref{sec:test}.

As specified in Section~\ref{sec: join and cc}, the cross-correlation is discretized as a vector with 50 linearly-separated radius bins (in radian). Under the assumption that the cross-correlation at each radius bin follows a univariate Gaussian distribution and measurements from different bins are dependent as they are monotone increasing functions, which makes the cross-correlation function multivariate Gaussian (verified in Section~\ref{sec:test}). The (log) likelihood function can be defined as:
\begin{linenomath*}
\begin{equation}\label{eq:4}
    \text{log}\mathcal{L}(f_{\text{astro}}|\boldsymbol{\psi}^{(k)}_{\text{g,v}})=-\frac{1}{2}
    (\boldsymbol{\psi}^{(k)}_{\text{g,v}}-\boldsymbol{\mu})\boldsymbol{\Sigma}^{-1}(\boldsymbol{{\psi}^{(k)}_{\text{g,v}}}-\boldsymbol{\mu})^T,
\end{equation}   
\end{linenomath*}
where $\boldsymbol{\mu}$ is the expected mean of the cross-correlation of a sample of combined astrophysical and atmospheric events with $f_{\text{astro}}$ and $\boldsymbol{\Sigma}$ is the associated covariance matrix of the cross-correlation functions. Since we only observed a single collection of neutrino events, lacking a clear analytical expression, we need to estimate the covariance matrix with Monte Carlo synthetic samples, which can be practically difficult as we are working with a high volume of data and a continuous parameter space. To this end, the expected mean is estimated with Equation \eqref{eq:approx}, and the covariance matrix is estimated with the covariance matrix of the cross-correlation between a set of 1000 Monte Carlo simulations of atmospheric-only events and the galaxy catalog, which only needs to be computed once.

Suppose we generate $n$ Monte Carlo simulations of atmospheric-only events in total, we denote the cross-correlation at $k$ between the $i$th simulated atmospheric-only events and the galaxy catalog as $\boldsymbol{\psi^{(k)}_{\text{g,atm,i}}}$. Then an unbiased estimate of the covariance matrix in Equation~\eqref{eq:4} is given as:
\begin{linenomath*}
\begin{equation}\label{eq:estimation_}
   \boldsymbol{\hat{\Sigma}}= \frac{1}{n-1}\sum_{i=1}^{n}(\boldsymbol{\psi^{(k)}_{\text{g,atm,i}}} - \boldsymbol{\overline{\psi}})(\boldsymbol{\psi^{(k)}_{\text{g,atm,i}}} - \boldsymbol{\overline{\psi}})^T,
\end{equation}  
\end{linenomath*}
where $\boldsymbol{\overline{\psi}}$ is the sample mean of the $n$ computed cross-correlations. The inverse is directly taken on the unbiased estimation of the covariance matrix above. For a specific $f_{\text{astro}}$ and $k$, to estimate the mean defined with Equation \eqref{eq:approx}, we first generate $n$ atmospheric-only ($f_{\text{astro}}=0$) and astrophysical-only events ($f_{\text{astro}}=1$).  The $\boldsymbol{\psi^{(k)}_{\text{g,atm}}}$ and $\boldsymbol{\psi^{(k)}_{\text{g,astro}}}$ are estimated with the sample mean of $n$ cross-correlations between the two groups of simulations with the galaxy catalog, respectively. Then the estimate for $\boldsymbol{\psi^{(k)}_{\text{g,v}}}(r)$ is obtained by directly plugging in $f_{\text{astro}}$ and the two sample means in Equation \eqref{eq:approx}.

The statistical significance of the measured cross-correlation against the null hypothesis, which is defined as the cross-correlation between the atmospheric-only events and the source catalog, can be quantified with a likelihood ratio test with test statistics defined as:
\begin{linenomath*}
\begin{equation}\label{eq:5}
    \text{TS}\equiv 2[\text{log}\mathcal{L}(\hat{f}_{\text{astro}}) - \text{log}\mathcal{L}({f}_{\text{astro}}=0)],
\end{equation}   
\end{linenomath*}
where $\hat{f}_{\text{astro}}$ is the maximum likelihood estimation for ${f}_{\text{astro}}$. By Wilks' Theorem \citep{d543aecb-cd73-36d5-9101-f08a74f8e8c6}, the above test statistic asymptotically follows a $\chi^2$ distribution with a degree of freedom (df) equals the number of free parameters in the likelihood function, which is the ${f}_{\text{astro}}$ in our case (df=1). 

\subsection{Verification of Key Assumptions}\label{sec:test}
Here we verify the key assumptions stated in Section \ref{sec:likelihood} with $k=8$ at a few selected ${f}_{\text{astro}}$ values. We choose $k=8$ as it optimizes the statistical power of the likelihood ratio test and is computationally efficient. More discussion regarding how to choose $k$ for an analysis is presented in Section \ref{sec:choice_k}. 

\begin{figure}
\includegraphics[width= 1 \linewidth] {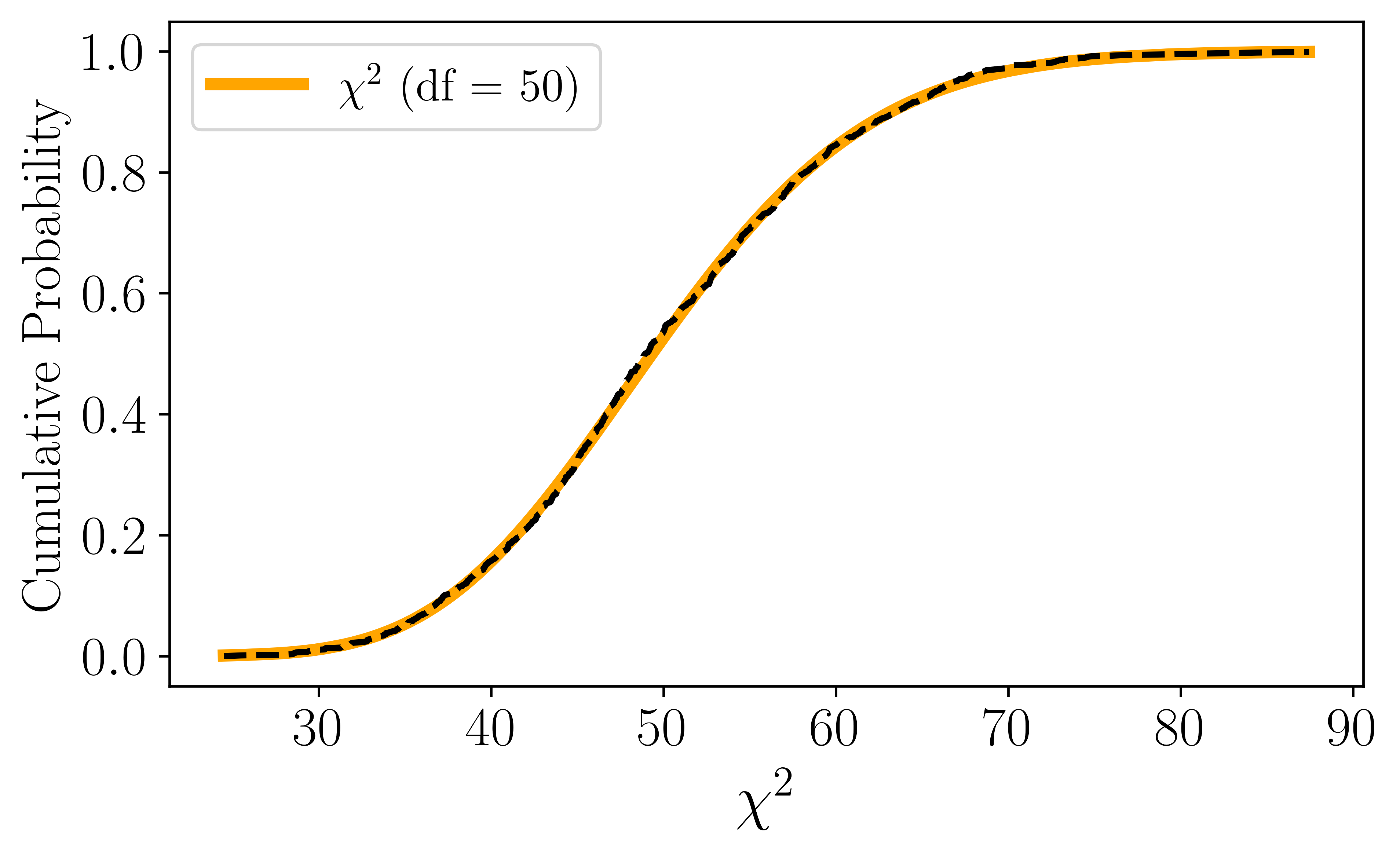} 
\caption{\label{fig:normality} The cumulative probabilities of the calculations from Equation \eqref{eq:6} with ${f}_{\text{astro}} = 0$ at $k=8$ (the black dashed curve) compared with that of the $\chi^2$ distribution with 50 degrees of freedom (orange solid curve).}
\end{figure}

Since we assume a multivariate Gaussian distribution of the estimated cross-correlation, we test the normality of the cross-correlations measured between the galaxy sample and 1000 synthetic background-only events (${f}_{\text{astro}}=0$) with a Henze-Zirkler multivariate normality test \citep{henze1990class}. Note that such tests can be performed using synthetic samples with different ${f}_{\text{astro}}$. The resulting p-value is $0.516$, so we do not reject the null hypothesis that the background-only cross-correlations follow a multivariate normal distribution. If the cross-correlations follow a multivariate normal distribution with mean $\boldsymbol{\mu}$ and covariance matrix $\boldsymbol{\Sigma}$, $\boldsymbol{\psi}^{(k)}_{\text{g,v}}\sim N_p(\boldsymbol{\mu}, \boldsymbol{\Sigma})$ and the covariance matrix $\boldsymbol{\Sigma}$ is positive definite, then we have:
\begin{linenomath*}
\begin{equation}\label{eq:6}
    (\boldsymbol{\psi}^{(k)}_{\text{g,v}}-\boldsymbol{\mu})^{T}\boldsymbol{\Sigma}^{-1}(\boldsymbol{\psi}^{(k)}_{\text{g,v}}-\boldsymbol{\mu})\sim\chi_{p}^2,
\end{equation}    
\end{linenomath*}
where $p$ is the number of radius bins we selected when estimating the cross-correlations. In our case, we choose $p=50$. We can visually examine the conclusion of Equation \eqref{eq:6} by plotting the empirical cumulative probabilities of the calculated chi-squared values against that of a $\chi^2$ distribution with $p=50$ degrees of freedom. We give the visual examination for $k=8$ and ${f}_{\text{astro}}=0$ in Figure \ref{fig:normality}, which shows a close fit of the empirical CDF with the expected one. 

\begin{figure}
\includegraphics[width= 1 \linewidth] {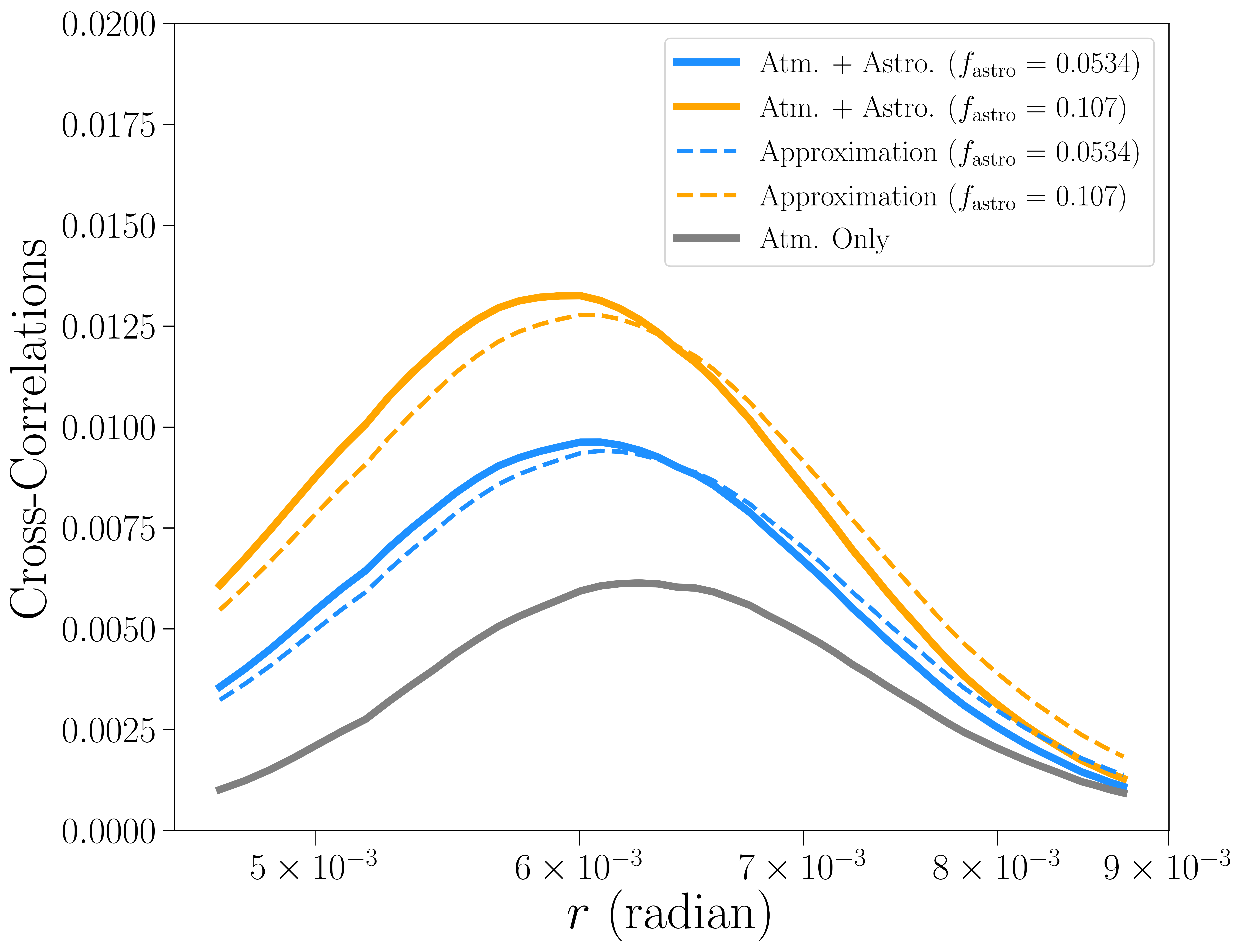} 
\caption{\label{fig:approx} Mean cross-correlations between 1000 synthetic neutrino samples with ${f}_{\text{astro}}=0.0534$ (blue lines) and $0.107$ (orange lines) and WISE-2MASS galaxy sample for $k=8$ together with the mean cross-correlations obtained from the background-only samples (the gray line). The approximations based on Equation \eqref{eq:approx} are shown as dashed curves.}
\end{figure}

Then we verify the approximation of Equation \eqref{eq:approx} with 1000 synthetic neutrino samples and background-only samples with ${f}_{\text{astro}}=0.0534$ and ${f}_{\text{astro}}=0.107$ for $k=8$ in Figure \ref{fig:approx}. Notice that the approximations are fairly good with slight right shifts, which may be introduced by the uncertainty when estimating $\boldsymbol{\psi^{(k)}_{\text{g,astro}}}(r)$ and $\boldsymbol{\psi^{(k)}_{\text{g,atm}}}(r)$. 

\subsection{Maximum Likelihood Estimation of \texorpdfstring{${f}_{\text{astro}}$}{}}\label{sec:MLE}
To evaluate the performance of the likelihood function from Equation \eqref{eq:4}, we compute the bias of ${f}_{\text{astro}}$, defined as the differences in its best-fit and true values, and show the results in Figure \ref{fig:MLE}. The blue error bars with diamonds indicate a likelihood using an approximate covariance matrix based on 1000 Monte Carlo simulations of atmospheric-only events. The orange error bars with squares indicates a likelihood with an accurate covariance matrix computed using the true ${f}_{\text{astro}}$. When using an accurate covariance matrix, we find no systematic bias for the estimation across a large range of ${f}_{\text{astro}}$ values. When using the approximate covariance matrix, we observe a systematic underestimation of ${f}_{\text{astro}}$. The differences between the estimations and true values are up to the 10\% level. Given the relatively small bias and the unknown ${f}_{\text{astro}}$, we adopt the likelihood with the approximate covariance matrix. In practice, this could be improved by maximizing the likelihood using tabulated covariance matrices for various ${f}_{\text{astro}}$ values and interpolating the tables during the estimation, rather than assuming a fixed covariance matrix with ${f}_{\text{astro}}=0$. 

\begin{figure}
\includegraphics[width= 1 \linewidth] {Fig_MLE_v2.png} 
\caption{\label{fig:MLE}Differences between the mean of the maximum likelihood estimations of the ${f}_{\text{astro}}$ of 1000 synthetics neutrino samples and the true ${f}_{\text{astro}}$. For each group of orange and blue error bars, the chosen ${f}_{\text{astro}}$ are, from left to right, 0, 0.0186, 0.0534, 0.107, 0.213, 0.320, 0.532. The error bars are based on the 2$\sigma$ of the distributions of the 1000 estimations. The blue error bars with diamonds shows the estimations where the covariance matrices are estimated with the background-only events. The orange error bars with squares} uses a likelihood with covariance matrices estimated using the true ${f}_{\text{astro}}$.
\end{figure}

\begin{figure}
\includegraphics[width= 1 \linewidth] {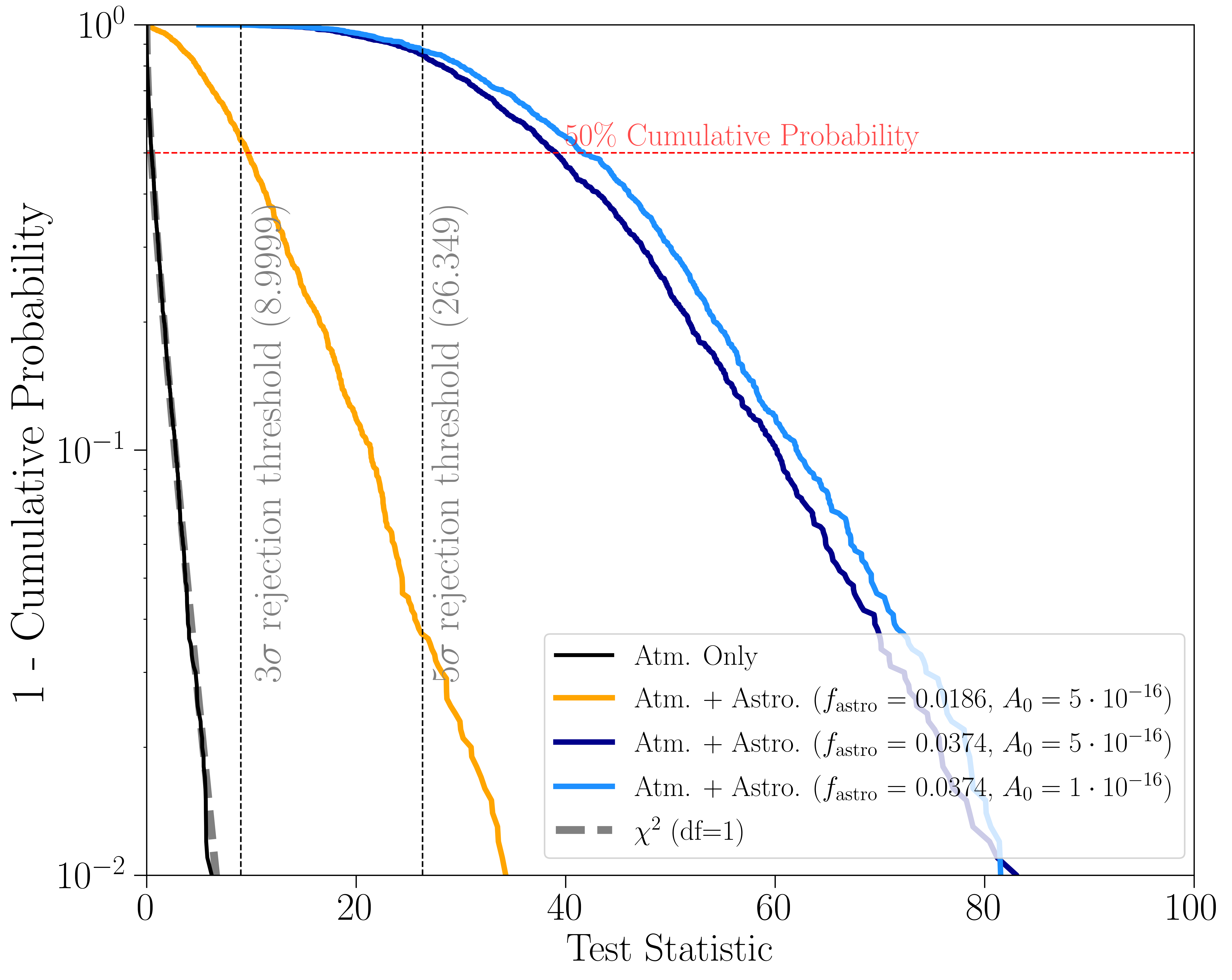} 
\caption{\label{fig:dp} Cumulative probabilities of test statistics of the cross-correlation between neutrino and galaxy samples via the KNN-CDF method. The black curve indicates the null hypothesis with ${f}_{\text{astro}}=0$, which is consistent with a $\chi^2$ distribution with 1 degree of freedom as indicated by the grey dashed curve. The orange curve indicates neutrino samples with ${f}_{\text{astro}}=1.86\%$ and $A_0 = 5\times 10^{-16}\,\rm GeV^{-1}cm^{-2}s^{-1}$ (about 5 neutrinos per source).  The blue curves are computed using neutrino samples with ${f}_{\text{astro}}=3.74\%$, with dark and light colors indicating $A_0 = 5$ and $1\times  10^{-16}\,\rm GeV^{-1}cm^{-2}s^{-1}$, respectively. The red dashed line indicates the 50\% cumulative probability.}
\end{figure}

\subsection{Constraints on Sensitivity and Discovery Potential}\label{sec:sensitivity}
Sensitivity and discovery potential are two crucial quantities used to evaluate the performance of the method. An ideal method should identify statistically significant cross-correlation signals with a relatively low ${f}_{\text{astro}}$. Following the IceCube convention (e.g., \citealp{IceCube:2019cia}), we define the sensitivity as the astrophysical neutrino flux, measured as ${f}_{\text{astro}}$ in a data sample, that yields a statistical power of 0.9 at a statistical significance level of $\alpha=0.5$ (0.5 false-positive probability). The discovery potential is defined as the astrophysical flux that results in a statistical power of 0.5 at a $5\sigma$ significance level ($\alpha=2.85\times10^{-7}$). To evaluate the performance of our framework, we compute the statistical power of the test at chosen ${f}_{\text{astro}}$ values with the test statistics defined in Equation \eqref{eq:5} under the hypothesis: 
\begin{linenomath*}
\begin{equation}
H_0:{f}_{\text{astro}}=0, \quad H_1:{f}_{\text{astro}}>0,
\end{equation}
\end{linenomath*}
where we choose $k=8$ for this calculation. The choice of $k$ depends on various factors, such as the data samples' densities and the events' angular uncertainties. We will discuss how to choose $k$ and its impact on the statistical power in Section \ref{sec:discussion}. The ${f}_{\text{astro}}$ values are chosen by varying the number of galaxies selected to inject synthetic neutrino events. The numbers we selected are 1500, 1750, 2000, 3500, 4000, 4250, 4500, 4750, and 5000, which covers the ${f}_{\text{astro}}$ from 1.86\% to 5.34\%.

Figure \ref{fig:dp} presents the cumulative probabilities of the test statistics obtained by cross-correlating the galaxy sample and synthetic neutrino events. We verify that the test statistic under the null hypothesis follows the $\chi^2$ distribution with 1 degree of freedom. Among the ${f}_{\text{astro}}$ values we tested, the case with ${f}_{\text{astro}}=1.86\%$ (corresponding to a total of 1750 galaxies selected to inject synthetic neutrino events) and ${f}_{\text{astro}}=3.74\%$ (corresponding to 3500 galaxies selected) already respectively reached the $3\sigma$ and $5\sigma$ significance levels with $>0.5$ statistical powers. We further tested another case with ${f}_{\text{astro}}=2.13\%$ (corresponding to a total of 2000 galaxies selected to inject synthetic neutrino events) and found that this failed to reach the $5\sigma$ significance levels with $>0.5$ statistical powers. As we evaluated the statistical power on a grid of $f_{\rm astro}$, the true discovery potential should be $[2.13\%, 3.74\%]$. A more accurate estimation can be obtained with a binary search on a more refined grid.  We find empirically that the statistical power of the test depends on the assumed source brightness. Under the same ${f}_{\text{astro}}$, the statistical power is higher with a lower $A_0$, which corresponds to a scenario with a larger population of dimmer individual sources (see Section \ref{sec:choice_k} for more discussion regarding $A_0$).

\section{Results}\label{sec:results}
\begin{figure}
\includegraphics[width= 1 \linewidth] {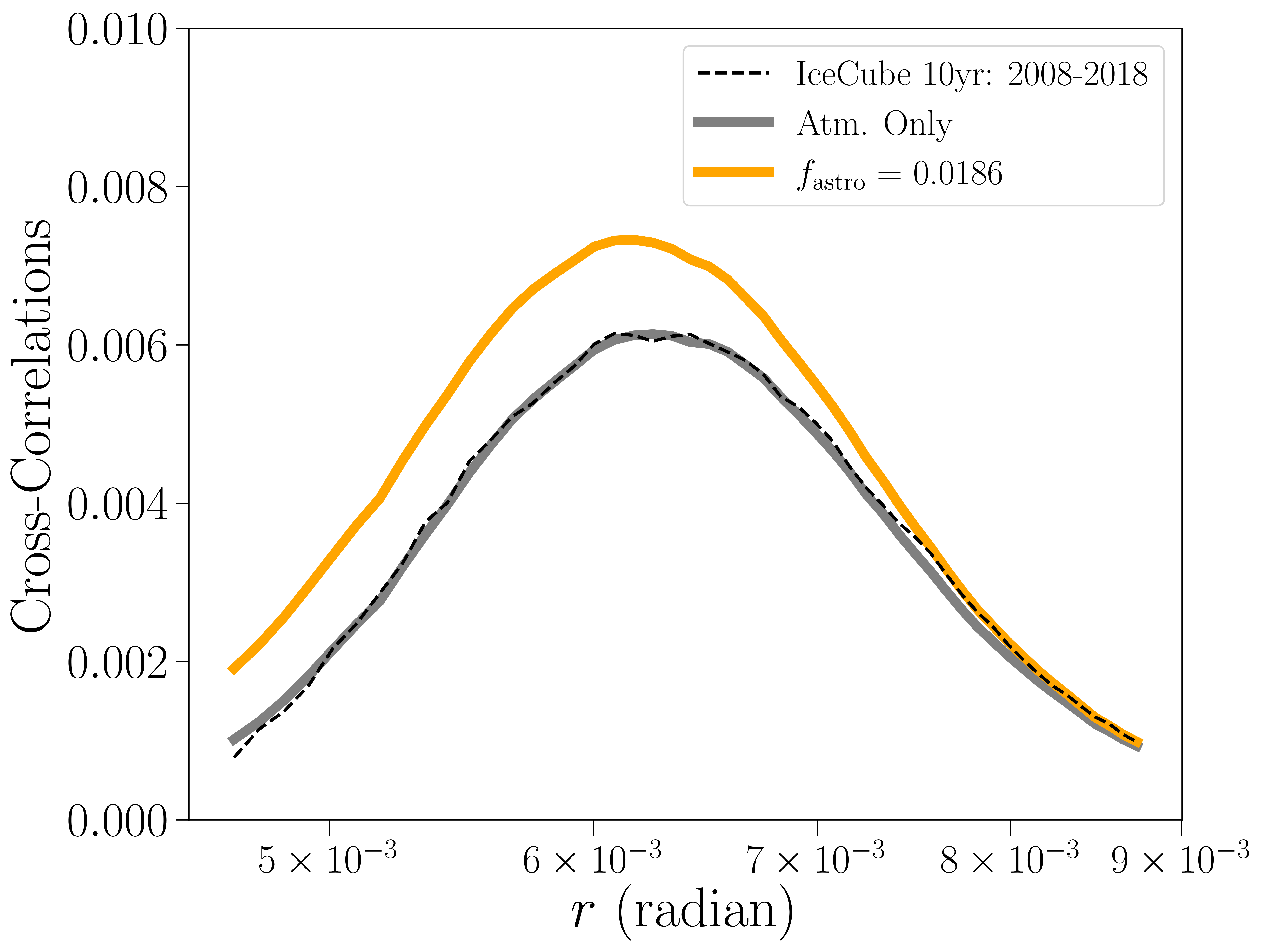} 
\caption{\label{fig:ic-cc} The cross-correlation between the IceCube ten-year point-source data and our down-sized WISE-2MASS galaxy sample across the scales of interest at $k=8$ (black dashed line). For comparison, we show the average cross-correlations from synthetic samples with ${f}_{\text{astro}}=0.0186$ (solid orange curve) and background-only events (solid gray curve).}
\end{figure}

Figure \ref{fig:ic-cc} presents the result obtained by applying our framework to the public IceCube ten-year point-source data \citep{abbasi2021icecube, publicData} and the down-sampled WISE-2MASS galaxy catalog. The dashed curve shows the $k$NN-CDFs cross-correlation across various angular sizes for $k=8$. It does not exhibit significant deviation from a cross-correlation using the background-only events. With the log-likelihood defined in Equation \eqref{eq:4}, we obtain a maximum likelihood best-fit $\hat{f}_{\text{astro}}=-0.00128$ with a likelihood ratio test statistic $\text{TS}=0.0589$ (corresponding to a p-value of $0.808$). 

We also take a Bayesian approach and sample the parameter space of ${f}_{\text{astro}}$ using Markov Chain Monte Carlo (MCMC) with the uniform prior of $f_{\rm astro} \in [-1,1]$. With 500 walkers and 2000 steps, the 90\% Bayesian credible interval based on the 5th and 95th quantiles region of the posterior is $(-0.0115, 0.009)$. The results suggest that we find no statistically significant evidence of astrophysical neutrinos in the IceCube ten-year point-source data that have spatial cross-correlations with the selected WISE–2MASS galaxy sample.

\section{Discussion and Conclusions
}\label{sec:discussion}
In this section, we discuss possible extensions and improvements of our methods as well as limitations. 

\subsection{The Choice of \texorpdfstring{$k$}{}}\label{sec:choice_k}
\begin{figure}
\includegraphics[width= 1 \linewidth] {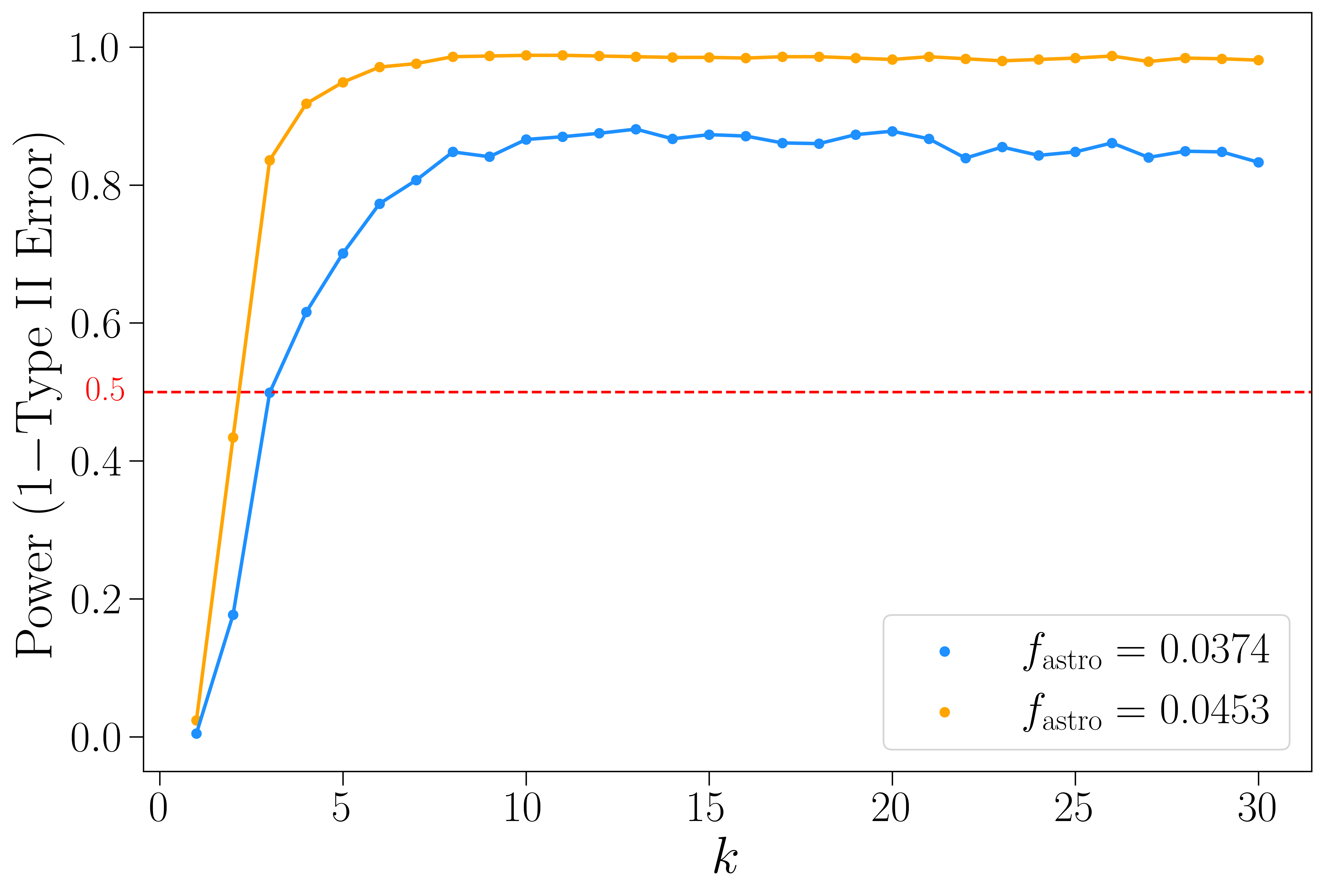} 
\caption{\label{fig:power} Statistical power (1 - Type II error) of the proposed likelihood-ratio tests, Equation~\eqref{eq:5}, on synthetic neutrino samples with ${f}_{\text{astro}}=0.0374$ (blue) and $0.0453$ (orange) across different $k$. The rejection threshold is set to be 26.349, which corresponds to a $5\sigma$ rejection threshold with a significance level of $\alpha=2.85\times10^{-7}$ when the null distribution is a $\chi^2$ distribution with 1 degree of freedom. A good choice of $k$ is the smallest one that maximizes the statistical power and (ideally) exceeds the 0.5 power threshold.}
\end{figure}

The $k$NN-CDFs are ways to measure the degree of spatial clustering at different scales, which are controlled by the choice of $k$. When computing the cross-correlations of two data samples, spatial similarities can occur at some specific scales. This indicates that we can expect to achieve the most significant cross-correlation signal (with higher power) at some specific $k$. To demonstrate such a proposition, we compute the statistical power of the likelihood ratio test formulated in Equation~\eqref{eq:5} of the cross-correlation between 1000 synthetic neutrino samples with ${f}_{\text{astro}}=0.0374 \text{ and } 0.0453$ and the WISE-2MASS galaxy sample under the $5\sigma$ rejection threshold. We plot the statistical power of the test with $k$ ranging from 1 to 30 in Figure \ref{fig:power}. 

At low $k$ values, the statistical power is low and reaches a plateau around $k=8$, and then becomes fairly stable.  Lower $k$ corresponds to the angular uncertainty ranges of the synthetic neutrino data which may lead to reduced statistical power, or there may simply not be significant correlated spatial clustering between the two samples at the lower scales. Such structures are successfully extracted by increasing the scale we are examining by increasing the $k$. While it is possible to compute at a higher $k$, it suffices to choose a lower $k$ that gives sufficiently high power to lower the computational cost. In practice, a reasonable approach for carrying out the analysis is to consider conducting similar experiments to select the $k$ by examining the relationship between the statistical power and $k$. An ideal choice of $k$ will be the smallest one that leads to a cross-correlation with $r$ that exceeds the angular uncertainties of the neutrino catalog and maximizes the statistical power and ideally exceeds the 0.5 power threshold.

\begin{figure}
\includegraphics[width=1 \linewidth] {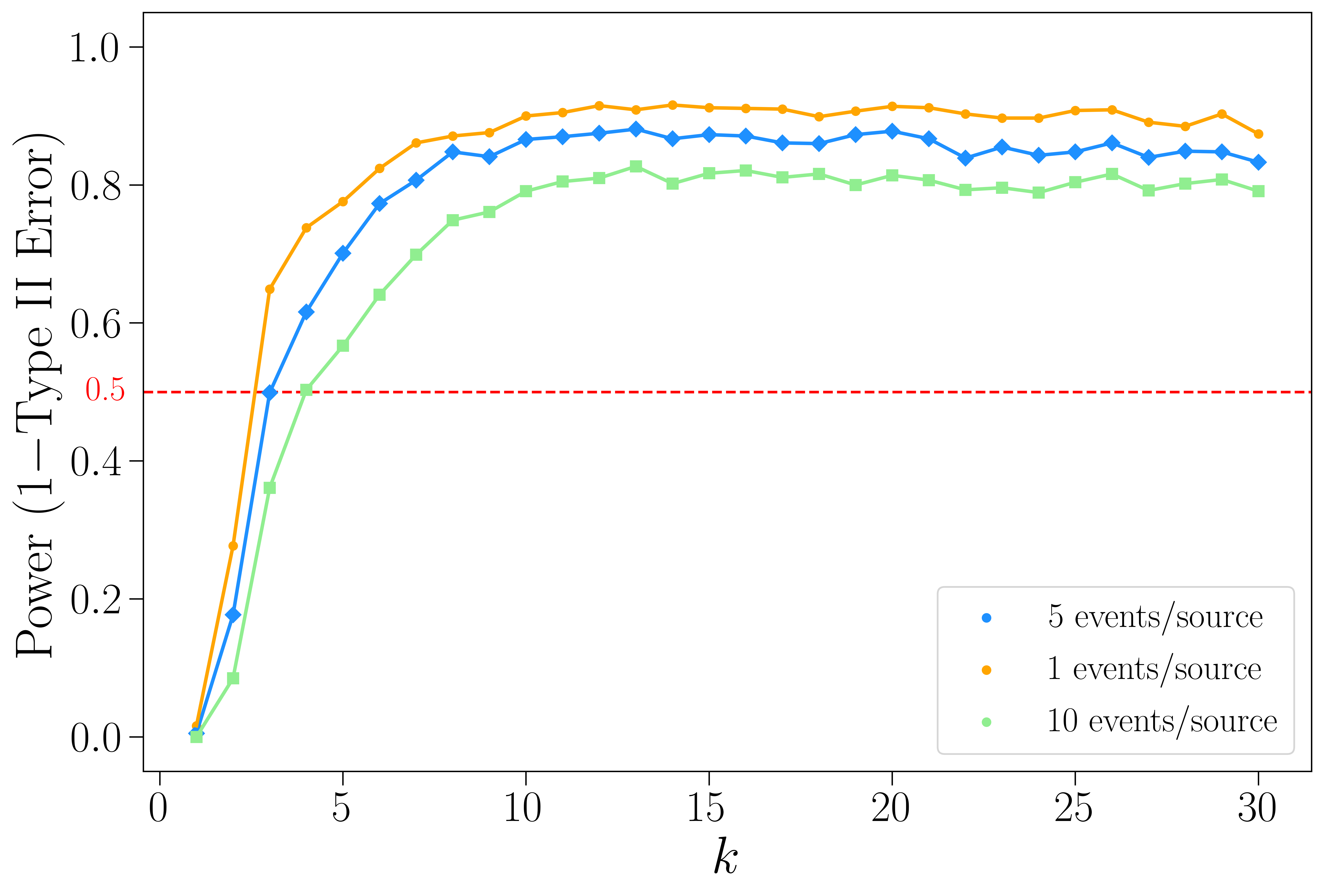} 
\caption{\label{fig: normalization} Statistical powers (1 - Type II error) of the proposed likelihood-ratio tests, Equation~\eqref{eq:5}, on synthetic neutrino samples with ${f}_{\text{astro}}=0.0374$ at three different brightness of the sources. The lines with orange circles, blue diamonds, and green squares correspond to a median of 1 event, 5 events, and 10 events per source, respectively.}
\end{figure}

As noted in Section~\ref{sec:sensitivity}, at the same ${f}_{\text{astro}}$, the statistical power of the test depends on the source brightness (number of events per galaxy).  Figure \ref{fig: normalization} compares the power of the test at ${f}_{\text{astro}}=0.0374$ under three source brightness assumptions: $A_0 = 1\times{10}^{-16}$, $5\times{10}^{-16}$, and $1\times{10}^{-15}\,\rm GeV^{-1}cm^{-2}s^{-1}$, which correspond to a median of 1, 5, and 10 events per source. The three cases present similar performances as a function of $k$, though a higher power is obtained for a lower $A_0$. This could be because the sources in the 1 event per source case better match the galaxy catalog. 

\subsection{Datasets with Different Sample Sizes}\label{sec:different_k}
In this paper, we have simplified the analysis by down-sampling the dataset with a larger sample size, the WISE-2MASS catalog, to match the sparser IceCube point-source data. Below we discuss how to implement an analysis with  $k_1\neq k_2$. 

One way to handle such a scenario is to adjust $k_1$ and $k_2$ based on the ratio of the sample sizes (average densities). This can put the two samples on equal footing when measuring their spatial similarities via the joint $k$NN-CDFs. Consider two homogeneous spatial Poisson processes  $D_1$ and $D_2$ with sample sizes $2n$ and $n$, respectively, in the same sample volume $V$. We expect the average $2$NN distances of the denser sample $D_1$ to be roughly equal to the average $1$NN distances of the sparser sample $D_2$. Indeed, in a two-dimensional Euclidean space, the actual ratio of average 2NN distance to 1NN distances is $\frac{3}{4}\sqrt{2}\approx1.0606$ (see Appendix \ref{appendix:ratio}). Thus an intuitive solution would be choosing $k_1=2$ for $D_1$ and $k_2=1$ for $D_2$. 
 
However, such a clear relationship may not exist for all $k$s. For example, the relationship discussed above for two homogeneous spatial Poisson processes does not exist for odd values of $k_1$. Also, if the sample sizes have a highly significant difference, it will be unrealistic to adjust the choices of $k$ simply based on the ratio of the sample sizes (average densities). Therefore, we propose a more efficient way for framing such problems by optimizing the areas between $\psi^{(k_1, k_2)}_{\text{g,astro}}(r)$ (the signal) and $\psi^{(k_1, k_2)}_{\text{g,atm}}$ (negative control) as defined in Section \ref{sec:likelihood}:
\begin{linenomath*}
 \begin{equation}\label{eq:9}
 (k_1,k_2) = \arg\max_{(k_1,k_2)}  \int_{0}^{r'} \psi^{(k_1, k_2)}_{\text{g,astro}}(r) -\psi^{(k_1, k_2)}_{\text{g,atm}}(r)\, dr. 
\end{equation}
\end{linenomath*}

We suggest choosing $r'$ to be slightly larger than the largest $k$NN distances measured across all selected $k$ for all data samples, which we adopted in this paper. Another approach can be selecting the range for integration based on the 5th and 95th quantiles of the measured $k$NN distances similar to what is outlined in Section \ref{sec: join and cc}. In practice, this can be optimized over experimental or synthetic data or with a train-test split on real data.

\begin{figure*}
\includegraphics[width= 0.9 \linewidth, scale=0.4] {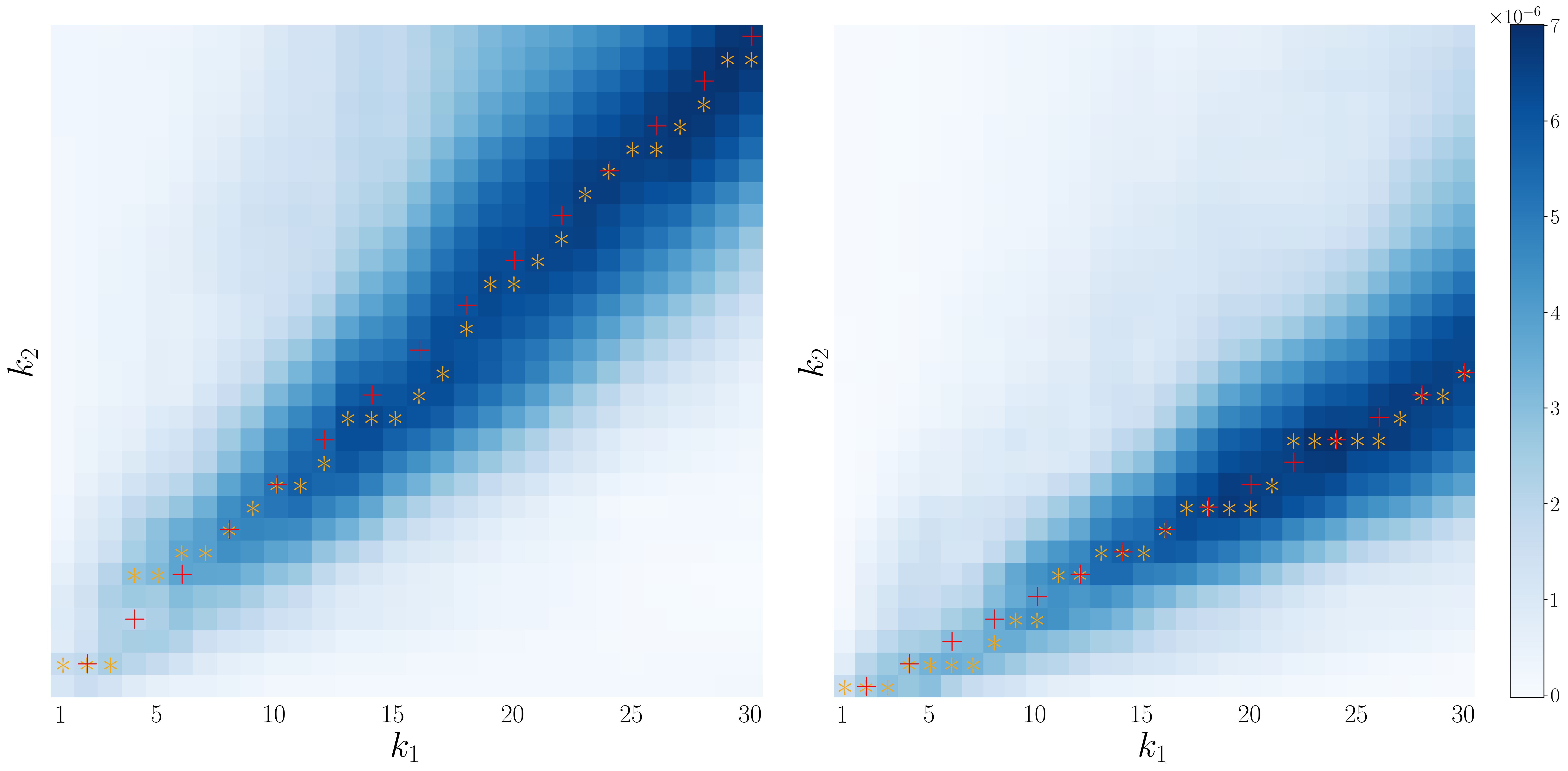} 
\centering
\caption{\label{fig:kk} Areas between the two cross-correlations calculated from the integral defined in Equation \eqref{eq:9} across all combinations of $k_1$ and $k_2$ up to $k_1=k_2=30$. $k_1$ (x-axis) corresponds to the neutrino sample and $k_2$ (y-axis) corresponds to the galaxy sample. The left panel demonstrates a case where the sample size of the galaxy sample matches the sample size of the neutrino sample, while the right panel illustrates a case where the sample size of the galaxy sample is roughly half of that of the neutrino sample. Orange stars stand for the $(k_1,k_2)$ pairs that maximize the integrals for a fixed $k_1$. The red pluses indicate the diagonal ($k_1=k_2$; left) and cells with $k_2 = k_1 / 2$ for all even values of $k_1$. 
%$(\frac{1}{2}k_1, k_1)$ .
}
\end{figure*}

We illustrate such an idea with a synthetic neutrino sample with ${f}_{\text{astro}}=0.0186$ and an atmospheric-only sample with ${f}_{\text{astro}}=0$, both with a sample size of $n=451,953$, in Figure \ref{fig:kk}. We first match the sample size of the WISE galaxy sample to both the synthetic neutrino sample and the atmospheric-only sample. Then we compute the integral defined in Equation \eqref{eq:9} across all combinations of $k_1$ and $k_2$ up to $k_1=k_2=30$. The integral results (area between the two cross-correlations) are displayed on the left panel of Figure \ref{fig:kk}. Notice that for a fixed $k_1$, the optimal choice of $k_2$ that gives the largest integrals (indicated by the orange stars) consistently occurs around the diagonal (indicated by the red pluses for even $k_1$). This matches our expectation and intuition that if the two samples have comparable or the same sample sizes, the maximum cross-correlation should occur (if there is such a correlation) when $k_1$ is roughly equal to $k_2$. 

Then, we down-sampled the WISE-2MASS galaxy sample so that its sample size is roughly half of the synthetic neutrino sample and atmospheric-only sample. We compute the same integral defined in Equation \eqref{eq:9} across all combinations of $k_1$ and $k_2$ up to $k_1=k_2=30$, and the results are displayed at the right panel of Figure \ref{fig:kk}. Notice that for a fixed $k_1$, the optimal choice of $k_2$ that gives the largest integrals (indicated by the orange stars) consistently occurs around $\frac{1}{2}k_1$ (denoted with the red pluses at entries with $(\frac{1}{2}k_1,k_2)$ for even $k_1$). This matches our discussions for the two homogeneous spatial Poisson processes with sample sizes $2n$ and $n$.

\begin{figure}
\centering
\includegraphics[width= 1 \linewidth] {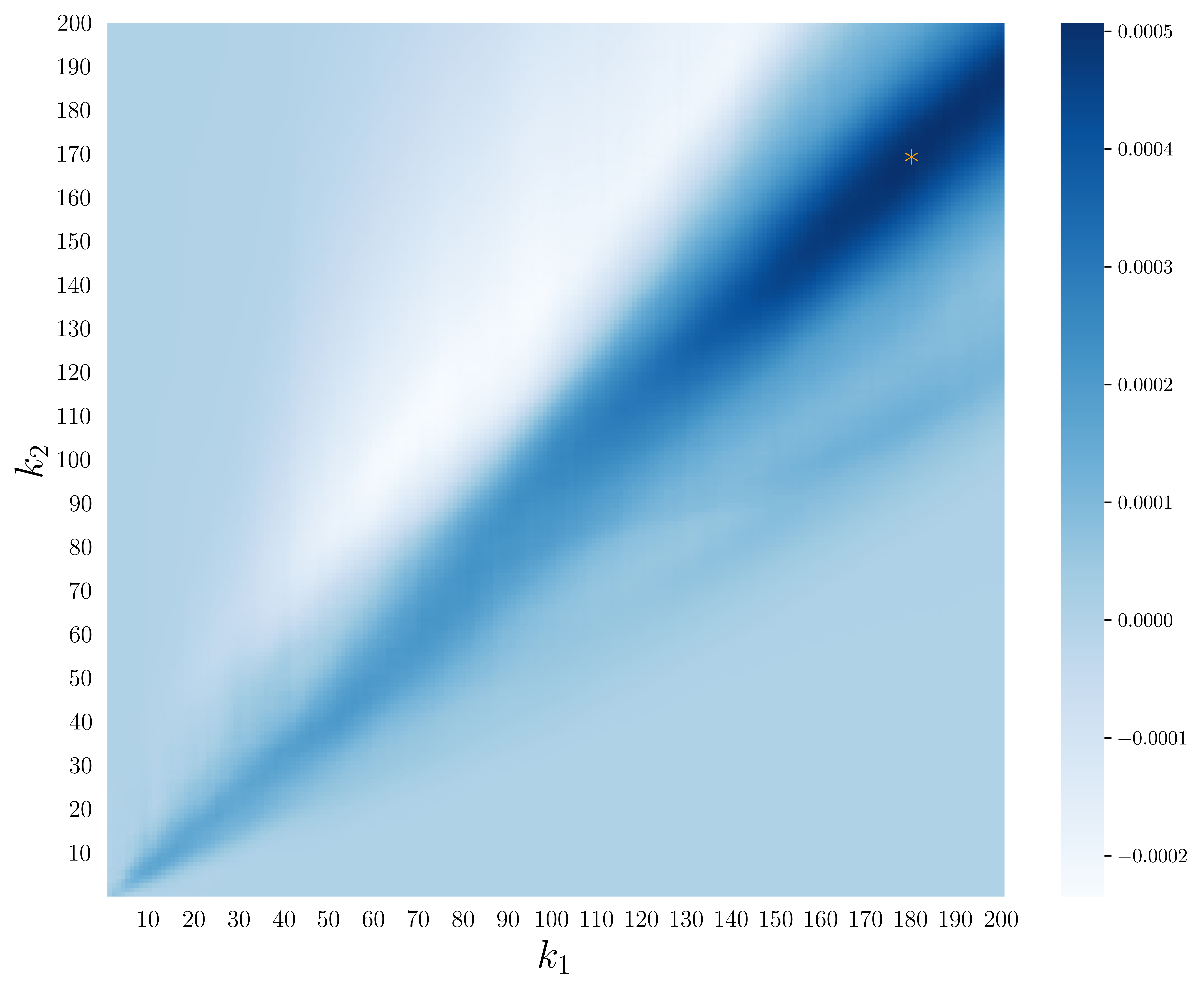} 
\caption{\label{fig:kk-small} Same as Figure~\ref{fig:kk} but with down-sampled neutrino data and a larger $k$ range. The orange star with $(k_1,k_2)=(180, 170)$ denotes the entry with maximum area across all combinations of $k_1$ and $k_2$. }
\end{figure}

On the other hand, if there is a significant correlation between the two samples, we expect to observe higher correlation values at certain scales rather than uniformly across all scales considered. Recall that $k$NN-CDFs can be understood as a measure of the degree of spatial clustering at different scales with ``probing spheres'' on the data sample. For both very small scale (smaller than the angular uncertainty of the observatory) and sufficiently large scale, the clustering of the data sample behaves more like background noise. For the equal-sample-size case, we do expect that the integral will increase to a maximum value and then start decreasing as $(k_1, k_2)$ increases.
The shape of the distribution of the integral defined in Equation \eqref{eq:9} is concave downward for both cases. We illustrate this observation in Figure \ref{fig:kk-small}, where we compute the same areas in the left panel of Figure \ref{fig:kk} but with neutrino data with a sample size of $n=10000$ and ${f}_{\text{astro}}=0.0186$. The decreased sample size enables us to observe the concave shape at a smaller $k$ with less computation time. Notice that the area peaks around the orange star (denoting the maximum area) with a concave downward shape. Especially for very small $(k_1,k_2)$, the areas become smaller and closer to zero. We also expect a decreasing trend of areas for $(k_1,k_2)$ that are larger than the orange star. In practice, we suggest considering both the optimization results and the power-$k$ curves as in Figure \ref{fig:dp} when selecting the optimal $k_1$ and $k_2$.

\subsection{Measurement with Higher-energy Events}\label{sec:highE}
\begin{figure}
\includegraphics[width= 1 \linewidth] {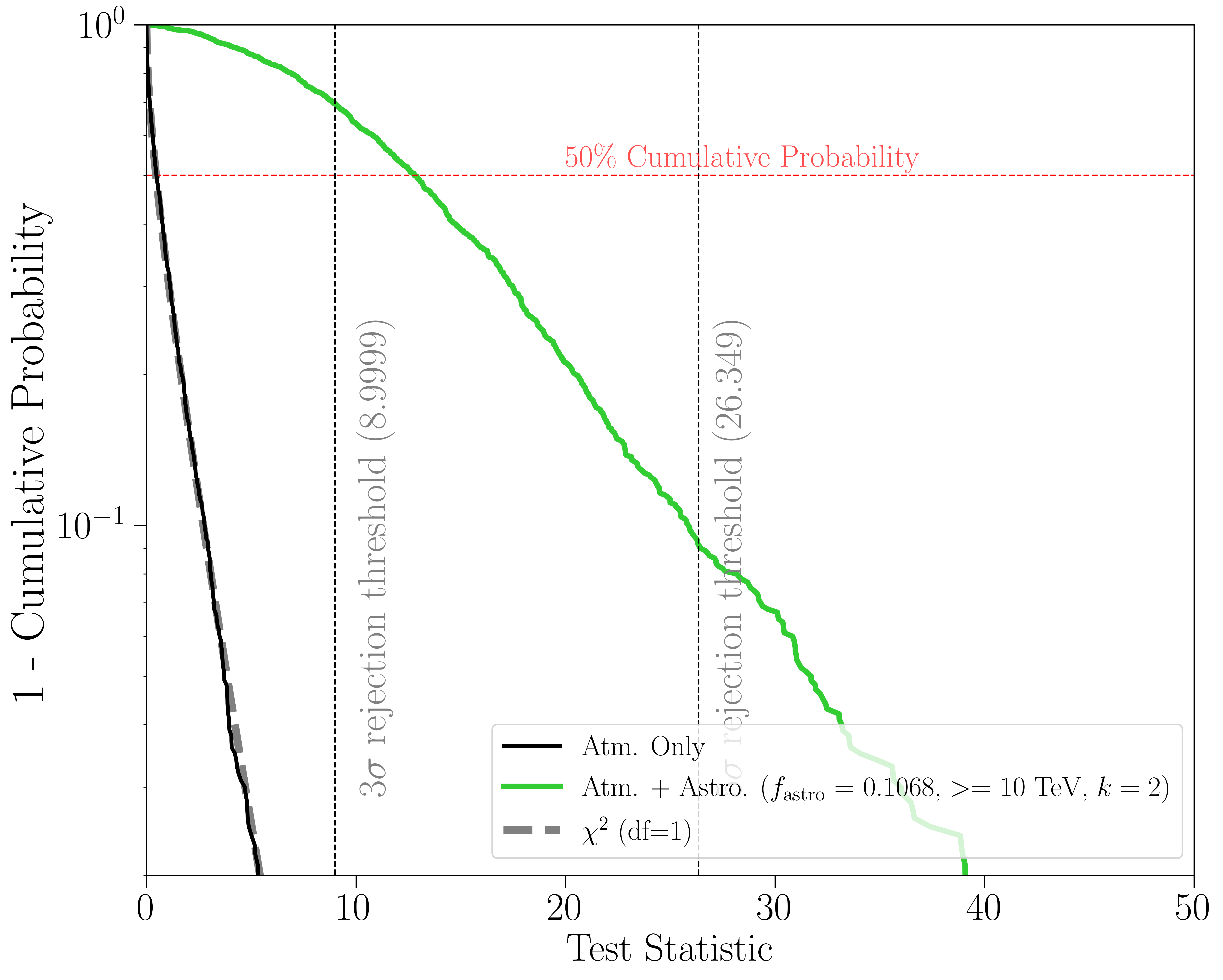} 
\caption{\label{fig:highe} Same as Figure \ref{fig:dp} but the related measurements are computed with neutrino events with energy $\geq 10^4 \, \rm GeV$. The green curve indicates neutrino samples with ${f}_{\text{astro}}=10.86\%$ and $A_0 = 5.8\times 10^{-16}\,\rm GeV^{-1}cm^{-2}s^{-1}$ (about 1 neutrino per source under the above energy setting).}
\end{figure}

We also tested our method's performance on neutrino samples with smaller sizes and higher purity of astrophysical events. This can be achieved by using neutrino events with higher energies. As a demonstration, we compute the $k$NN-CDFs cross-correlations using a neutrino sample selected from the public ten-year point source data with energy higher than $10^4 \, \rm GeV$. The sample size is reduced to $n=1539$ and we adopt a smaller $k$ for the analysis, $k=2$. Figure \ref{fig:highe} presents the sensitivity of such an analysis. Similar to the procedures as described in Section \ref{sec:sensitivity}, we tested several ${f}_{\text{astro}}$ values and found that ${f}_{\text{astro}}=10.68\%$ is needed to reach the $3\sigma$ significance level with $>0.5$ statistical power. 

\subsection{Conclusion and Future Work}\label{sec:conclusion}
Compared with previous cross-correlation studies, our framework can better handle the non-Gaussian features in the dataset with improved computational efficiency and sensitivity. The $k$NN-CDFs and the associated cross-correlations also provide an interpretable characterization of the spatial clustering and spatial similarities between data samples. While there are many different functions for summarizing spatial point patterns, the $k$NN-CDF provides extra flexibility due to the selection of $k$, which is one of the reasons we consider this approach. This method is also a generalization of studies using the two-point correlation function such as \citet{Fang2020}. Such methodologies can lead to a better understanding of the spatial structure of the detected neutrinos across different scales.

We find no significant cross-correlation between the IceCube public ten-year point-source data and the selected WISE-2MASS galaxy sample. This is not surprising as our analysis does not account for the energy information of the neutrinos. With the majority of the neutrino events being atmospheric, the $f_{\rm astro}$ of the sample is well below the percent-level, which is needed to achieve a detection as shown in Figure~\ref{fig:dp}. 

An improvement to our study would be to invoke an energy-dependent analysis, for example by performing the cross-correlation study in different neutrino energy bins \citep{2023arXiv230803978G} or using neutrino samples with higher purity \citep{IceCube:2020wum,Abbasi:2024jro}. Our method currently works under the assumption that the neutrino and galaxy catalogs have roughly the same sample sizes on the spherical surface. As discussed in section \ref{sec:different_k}, an important future direction is to generalize the framework to catalogs with significantly different sample sizes. An alternative way to solve such problems can be approximating the denser galaxy catalog as a continuous field on some HEALPix pixelation, and then apply the $k$NN version of the cross-correlation between discrete tracers and continuous fields as formulated in \citet{banerjee2023tracer}. This potentially retains the small-scale information in the galaxies while avoiding down-sampling the catalog. Such a formalism has been applied to study gravitational wave (GW) transient sources and galaxy samples \citep{gupta2024spatial}. 

Another improvement of the analysis would be implementing more realistic synthetic astrophysical events generation. Currently, as described in Section \ref{sec:synthetic}, the sources are assumed to be equally bright with the same $A_0$. The number of events emitted by a source varies as a Poisson random number with the mean being the assumed source flux integrated over the detector effective area for a given source direction. This could lead to a biased estimation of the method's sensitivity. A more realistic setup,  for example, may weight $A_0$ by the inverse square of the galaxies' luminosity distances. However, as our method currently works under the assumption that the neutrino and galaxy catalogs have roughly the same sample sizes, it is difficult to weight $A_0$ while keeping the sample sizes consistent. This could be made possible after generalizing the framework to catalogs with significantly different sample sizes. 
%Fixing $A_0$ helps to more easily control the astrophysical events generated from the galaxies and ${f}_{\text{astro}}$. Adjusting $A_0$ based on the luminosity distances while controlling the sample size and ${f}_{\text{astro}}$ can be computationally cumbersome. Therefore, generalizing the framework to catalogs with significantly different sample sizes can also make the above implementation more efficient.}

Also, we only select a specific $k$ for the statistical testing and estimation. It is possible to extend the framework by incorporating multiple $k$NN-CDFs in a single analysis during the statistical inference. Other possible approaches to analyze the neutrino point-source data are to consider other spatial point process functional summaries using the proposed framework. Another interesting approach is to consider the data as marked point processes, where the marks on the point processes are a label of either ``IceCube'' or ``WISE'', and then statistical tests can be applied to check for independence between the marks. \footnote{We thank an editor for this suggestion.} The study may be extended by using cosmological survey products with higher densities, different redshift ranges, and larger sky coverage \citep{Fang2020}. Different tracers other than galaxies can be analyzed as well. 

\software{The code for the $k$NN-CDFs, joint $k$NN-CDFs, and $k$NN-CDFs Cross-correlation computation and synthetic event generation is publicly available at \url{https://github.com/Zhuoyang666/IceCubekNN} and the Zenodo repository at \url{https://zenodo.org/records/13352530} \citep{zhou_2024_13352530}. {\it Jupyter Notebook} files and code for reproducing some of the key figures are in the paper also provided. All code is written in {\it Python}.}

\begin{acknowledgments}
We thank the reviewer and editors for helpful suggestions.
We thank Michael Larson for his help with using the IceCube public ten-year point-source data. 
This research was performed using the computing resources and assistance of the University of Wisconsin-Madison Center For High Throughput Computing (CHTC) in the Department of Computer Sciences \citep{chtc}. The CHTC is supported by the University of Wisconsin-Madison, the Advanced Computing Initiative, the Wisconsin Alumni Research Foundation, the Wisconsin Institutes for Discovery, and the National Science Foundation, and is an active member of the Open Science Grid, which is supported by the National Science Foundation and the U.S. Department of Energy's Office of Science. 
ZZ and JCK acknowledge support from NSF under Grant Number DMS 2038556. ZZ acknowledges support from the Hilldale Undergraduate/Faculty Research Fellowship awarded by the University of Wisconsin-Madison.
The work of KF is supported by the Office of the Vice Chancellor for Research and Graduate Education at the University of Wisconsin-Madison with funding from the Wisconsin Alumni Research Foundation. KF acknowledges support from the National Science Foundation (PHY-2110821, PHY-2238916) and NASA (NMH211ZDA001N-{\it Fermi}). This work was supported by a grant from the Simons Foundation (00001470, KF). KF acknowledges the support of the Sloan Research Fellowship.

\end{acknowledgments}

\appendix
\restartappendixnumbering
\section{Average 1NN and 2NN Distances for Homogeneous Spatial Poisson Processes}\label{appendix:ratio}
The cumulative density function of the nearest neighbor (1NN) distance for homogeneous spatial Poisson processes is written as \citet{Banerjee2021a}:
\begin{linenomath*}
\begin{equation}
    \text{CDF}_{\text{1NN}}(V)=1-e^{-\lambda V},
\end{equation}    
\end{linenomath*}
where $V=\pi r^2$ and $\lambda$ is the intensity of the homogeneous spatial Poisson process. Thus, we have:
\begin{linenomath*}
\begin{equation}
    F_{R}(r)_{\text{1NN}}= \text{P}(R\leq r)= 1-e^{-\lambda \pi r^2}.
\end{equation}    
\end{linenomath*}
The probability mass function can be obtained by differentiating the cumulative density function above:
\begin{linenomath*}
\begin{equation}
    f(r)_{\text{1NN}}=\frac{d}{d r}F_{R}(r)_{\text{1NN}}=2\lambda\pi r\cdot e^{-\lambda\pi r^2}.
\end{equation}   
\end{linenomath*}
Then the expected 1NN distance can be obtained by integrating the above mass function on the real line as:
\begin{linenomath*}
\begin{equation}
    E[R]_{\text{1NN}}=\int_{}^{} r \cdot2\lambda\pi r\cdot e^{-\lambda\pi r^2}\, dr = \frac{1}{2\sqrt{\lambda}}.
\end{equation}    
\end{linenomath*}

Similarly, with $V=\pi r^2$ and intensity as $\lambda$, the cumulative distribution function of the 2NN distances for homogeneous spatial Poisson processes can be written as \citet{Banerjee2021a}:
\begin{linenomath*}
\begin{equation}
    F_{R}(r)_{\text{2NN}}= \text{P}(R\leq r)= 1-e^{-\lambda \pi r^2} - (\lambda \pi r^2)e^{-\lambda \pi r^2}.
\end{equation}  
\end{linenomath*}
The probability mass function can be obtained by differentiating the cumulative density function above:
\begin{linenomath*}
\begin{equation}
    f(r)_{\text{2NN}}=\frac{d}{d r}F_{R}(r)_{\text{2NN}}=2\lambda^2\pi^2 r^3\cdot e^{-\lambda\pi r^2}.
\end{equation}    
\end{linenomath*}
Then the expected 2NN distance can be obtained by integrating the above mass function:
\begin{linenomath*}
\begin{equation}
    E[R]_{\text{2NN}}=\int_{}^{} r \cdot 2\lambda^2\pi^2 r^3\cdot e^{-\lambda\pi r^2}\, dr = \frac{3}{4\sqrt{\lambda}}.
\end{equation}    
\end{linenomath*}

Consider two homogeneous spatial Poisson processes, $X_1$ and $X_2$, with sample sizes $n$ and $2n$, respectively, on a two-dimensional surface. Then we have intensities $\lambda_{X_1}=\frac{1}{2}\lambda_{X_2}$, and the average 1NN distance of $X_1$ and average 2NN distance of $X_2$ can be written as:
\begin{linenomath*}
\begin{equation}
    E[R_{X_1}]_{\text{1NN}}=\frac{1}{2\sqrt{\lambda_{X_1}}},
\end{equation}    
\begin{equation}
  E[R_{X_2}]_{\text{2NN}}=\frac{3}{4\sqrt{2\lambda_{X_1}}}.
\end{equation}    
\end{linenomath*}
The ratio of the 2NN distance of $X_2$ to the average 1NN distance of $X_1$ is then $\dfrac{E[R_{X_2}]_{\text{2NN}}}{E[R_{X_1}]_{\text{1NN}} }=\frac{3}{4}\sqrt{2}\approx1.0606$.

\bibliographystyle{aasjournal}
\bibliography{ref} 

\end{CJK*}
\end{document}